\documentstyle[twocolumn,prb,aps,floats,epsf]{revtex}
\begin{document}
\draft

\twocolumn[\hsize\textwidth\columnwidth\hsize\csname
@twocolumnfalse\endcsname

\title{Structure and oxidation kinetics of the Si(100)-SiO$_2$ interface}
\author{Kwok-On Ng and David Vanderbilt}
\address{Department of Physics and Astronomy,
  Rutgers University, Piscataway, NJ 08855-0849}

\date{October 28, 1988}

\maketitle

\begin{abstract}

We present first-principles calculations of the structural and
electronic properties of Si(001)-SiO$_2$ interfaces.
We first arrive at reasonable structures for the c-Si/a-SiO$_2$
interface via a Monte-Carlo simulated annealing applied to
an empirical interatomic potential, and then relax these
structures using first-principles calculations within the
framework of density-functional theory.
We find a transition region at the interface, having a thickness on
the order of 20\AA, in which there is some oxygen deficiency and
a corresponding presence of sub-oxide Si species (mostly
Si$^{+2}$ and Si$^{+3}$).  Distributions of bond lengths and
bond angles, and the nature of the electronic states at the
interface, are investigated and discussed.
The behavior of atomic oxygen in a-SiO$_2$ is also investigated.
The peroxyl linkage configuration is found to be lower in
energy than interstitial or threefold configurations.  Based on
these results, we suggest a possible mechanism for
oxygen diffusion in a-SiO$_2$ that may be relevant to the
oxidation process.

\end{abstract}

\vskip2pc]

\narrowtext

\section{Introduction}

Understanding the atomic structure of ultrathin silicon dioxide films on
Si(100) substrates is an outstanding problem  of great importance for
microelectronic applications.\cite{helms94,si-device} The thickness of
these films in commercial devices has dropped well below 100\AA. Film
thicknesses of 40\AA\ and below are now being explored in experimental
devices. Although such ultrathin oxide films are now in wide use,
and considerable information about them has been amassed using a
variety of different experimental techniques,
\cite{helms94,physics-sio2,atkinson85,si-system,chem-sio2}
there is little consensus about the microscopic structure and oxidation
kinetics of the films.  Among the factors that have so far hindered
detailed studies of the interface are
the amorphous nature of the SiO$_2$ region, the difficulty of probing a
buried interface, and the dependence of sample properties on
preparation conditions. As a result, several issues relating to the
detailed chemical structure of the interface, and to the oxidation mechanism
of thin oxide films, continue to be controversial.

Photoemission spectroscopy (PES) and photoelectron spectroscopy (XPS)
have been the major experimental techniques used to study the
interface structure. These experiments indicate the presence of
a transition region of sub-oxidized Si near the interface, with all
three intermediate partial oxidation states observed. However, different
experiments suggest conflicting pictures of the structure. For example, PES
experiments indicate that the total suboxide is about two monolayers,
\cite{hollinger,himpsel} while XPS results suggest it is one monolayer
thick. \cite{grunthaner87}  Using medium-energy ion-scattering spectroscopy
(MEIS) to study the interface, Gusev {\it et
al.}\cite{gusev95} and Lu {\it et al.}\cite{lu-apl} have
studied sequential isotopic exposures of oxygen and found that the
growth mechanism of interfaces depends on the thickness of the films.
Further complicating the task has been the
lack of suitable interface models which serve as reference points for
analysis of spectroscopic results. Theoretical works by Pasquarello
{\it et al.}\cite{pas-prl,pas-apl}
have attempted to construct interface models by attaching
tridymite, a crystalline form of SiO$_2$, to Si(100). Although
different suboxide states can be created, one
would expect such an artificial model would introduce significant
stress across the interface. A recent experimental work suggest
an abrupt interface model,\cite{luh97} as opposed to the graded
interface suggested from earlier
work.\cite{himpsel,grunthaner87} Additional uncertainty has been
generated by some other studies about the interpretation of the
sub-oxide peaks observed.\cite{holl93,holl94}

Another aspect that is of great interest is the oxidation
kinetics for thin films. The formation of thick oxide
films ($>$ 200\AA) is well described phenomenologically by the
Deal-Grove model,\cite{deal-grove} in which molecular oxygen
diffuses to the SiO$_2$/Si
interface and reacts with silicon at the interface. This model predicts
a linear relationship between oxide thickness and oxidation time. However,
it is well known that the Deal-Grove model breaks down
for the case of ultrathin films ($<$100\AA); in this case, the
oxidation kinetics have been shown to be faster
than would be expected from the linear
relationship.\cite{si-system,gusev95,irene88,mott89}
However, there is little consensus about the mechanisms of oxidation
kinetics in this case.  Several phenomenological models have been
proposed.\cite{mott89,sofield95,stoneham87}
Some of them fit the experimental data on
oxidation kinetics quite satisfactorily with a large number of fitting
parameters. However, most of the models are without direct
experimental support. An analysis of kinetic results alone does not
allow one to distinguish among different models. A few key problems
associated with the oxidation mechanism still remain unresolved.
For example, it
is uncertain exactly where oxidation takes place, and the
mechanism of oxygen diffusion through a-SiO$_2$ region in not clear.

In this paper, we choose to address two main issues, namely, the
microscopic structure of Si/SiO$_2$ near the interface and the
behavior of atomic oxygen in a-SiO$_2$. Firstly, we arrive at
plausible structures for the Si/SiO$_2$ interface by performing
Monte Carlo simulated annealing using an empirical potential.
Then we relax these structures and study their properties via
first-principles density-functional calculations.
We obtain the statistical distribution of
sub-oxide species near the interface region, and find a graded
transition region with a width of about 10 \AA. Secondly, for
excess atomic oxygen in a-SiO$_2$, we compare
the energy of the interstitial
and peroxyl bridge configurations. It is found that
the latter is always lower in energy.
This suggests a possible mechanism of
oxygen hopping among neighboring Si--Si bonds.

The plan of the paper is as follows. Section II gives a brief
description of the technique used to perform the calculations. In
Sec.~III we present our work on the microscopic structure of
the Si/SiO$_2$ interface, and Sec.~IV contains the results on oxidation
kinetics of atomic oxygen in a-SiO$_2$.  We summarize the work
in Sec.~V.

\section{Methods}

Our theoretical analysis consists mainly of two parts. In the first
part, we employ Metropolis Monte Carlo (MC) simulations together with an
empirical model potential representing the structural energies to
arrive, via simulated annealing, at candidate structures for the
c-Si/a-SiO$_2$ interface.  In the second part of the analysis,
these candidate structures are relaxed and analyzed using
first-principles plane-wave pseudopotential calculations.
Because the structures of interest involve coordination changes, we
have followed Hamann\cite{hamann96} in employing a generalized
gradient approximation (GGA) to the exchange-correlation
potential, specifically that of Perdew, Burke and Ernzerhof
(PBE).\cite{pbe}

For use in generating the c-Si/a-SiO$_2$ interface structures within
the MC approach, we designed a simple empirical-potential
model for the structural energy that is roughly based on
the formalism of Keating.\cite{keating}  The only degrees of freedom
that appear in this model are the Si atom coordinates; a direct
Si--Si bond is designated as a ``short'' (S) bond, while a Si--O--Si
bridge configuration is designated as a ``long'' (L) bond. In this way,
the whole c-Si/a-SiO$_2$ structure is replaced by a network of
``short'' and ``long'' bonds between Si atoms.  Its energy is
taken to be
\begin{equation}
E = {1\over2} \sum_i k_r^{(i)} (d_i-d_0^{(i)})^2 + {1\over2}
\sum_{i\neq j} k_\theta^{(ij)} (c_{ij}-c_0^{(ij)})^2 
\;\;.
\end{equation}
The first term represents the bond stretching term, where $k_r^{(i)}$
and $d_0^{(i)}$ take values
$k_r^{\rm S} = 9.08$eV/\AA$^2$ and $d_0^{\rm S} = 2.35$\AA\ for a
``short'' bond, or
$k_r^{\rm L} = 1.89$eV/\AA$^2$ and $d_0^{\rm L} = 3.04$\AA\ for a
``long'' bond.
Similarly, the second term represents the bond-bending term, where
$k_\theta^{(ij)}$ takes values
$k_\theta^{\rm SS} = 3.58$eV,
$k_\theta^{\rm SL} = 3.81$eV, and
$k_\theta^{\rm LL} = 4.03$eV, where $i,j$ are summed over neighboring
bonds only. Here $c_{ij}$ is the cosine of the angle subtended by bonds $i$
and $j$, and $c_0^{(ij)}=\cos(109.47^\circ)= -1/3$ is the corresponding
reference value.
The parameters for the empirical
potential were chosen such that they reproduce not only the
equilibrium structures, but also the bulk and shear moduli, of
crystalline Si and $\beta$-cristobalite structures.
The only parameter that cannot be fixed
empirically is the bond-bending term $k_\theta^{\rm SL}$ associated
with a mixture of S and L bonds. However, it is found that
$k_\theta^{\rm SS}$ and $k_\theta^{\rm LL}$ are very close to each
other, so $k_\theta^{\rm SL}$ was chosen as the average of these. 

\begin{figure}
\epsfxsize=1.7in
\centerline{\epsfbox{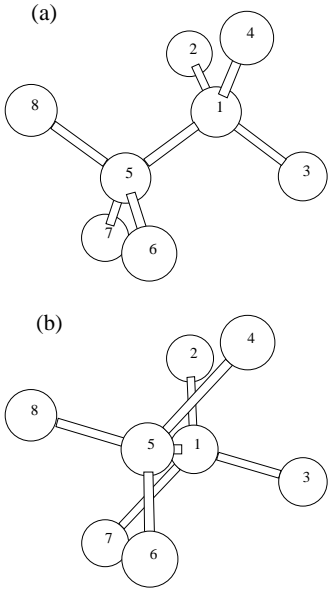}}
\vskip 0.5cm
\caption{The WWW bond-switching move (Ref.~\protect\onlinecite{www})
used in the Monte Carlo simulations.
(a) Initial configuration.  Atoms 1 and 4 are first
neighbors, and atoms 1 and 7 are second neighbors.
(b) After bond switch.  Atoms 1 and 7 are now first
neighbors, while atoms 1 and 4 are second neighbors.}
\label{fig:www}
\end{figure}

There are two types of discrete moves allowed in our MC simulations,
namely, bond switching and bond conversion moves. For the bond
switching moves, we
adopted the algorithm of Wooten, Winer, and Weaire (WWW).\cite{www} To
randomize the structure, one of the second neighbors of a randomly
chosen  atom is
switched to become its first neighbor, and that first neighbor becomes
the second neighbor (see Fig.\ \ref{fig:www}). Notice that the bonds being
switched can be either S or L bonds. For the bond
conversion moves, a randomly chosen S or L bond is swapped with a
neighboring bond of opposite type. In all of the discrete moves, the
total number of S and L bonds are conserved, meaning that the number of
oxygen atoms in the system does not change during the MC
simulations. After each move, the structure is relaxed within the
Keating-like model, and the energy difference between the relaxed structures
before and after the move is computed. The Metropolis algorithm is then
employed to decide whether the new structure is to be accepted. In the case
of the Si/SiO$_2$ interface, certain atoms in the c-Si region are
fixed (i.e., not allowed to relax, and no bond switching or conversion
moves involving them are allowed).

After generating structures from the MC simulations,
first-principles plane-wave pseudopotential calculations are carried
out to produce accurate and relaxed structures. The exchange and
correlation energy and potential are implemented in the PBE-GGA
scheme.\cite{pbe}
The pseudopotential for O is that used in Ref.\ \onlinecite{madhav},
and was generated using an ultrasoft pseudopotential
scheme.\cite{vand}
The ultrasoft pseudopotential for Si is generated with a core
radius of 1.3 a.u. and valence states of $3s$ and $3p$.
Self-consistent total-energy and force calculations are
used to relax the atomic coordinates until the forces are less than
0.25 eV/\AA.
A cutoff energy of 25 Ry is used for the calculations in Sec.~IV,
but this was reduced to 15 Ry for the majority of total-energy
calculations of interface structures presented in Sec.~III.
The low energy cutoff for the interface structures was
chosen so as to reduce the computational burden. It is found that when
we raise the cutoff energy to 20 Ry, there is only a slight effect on
the structural configurations. More specifically, there is less
than $1\%$ of error in bond lengths and $5\%$ of error in bond
angles when we compare
structures generated with different cutoff energies.
So, a cutoff
energy of 15 Ry is enough to provide accurate results for our
analysis. For all cases, the Brillouin zone sampling consists of
a single k-point at $\Gamma$. This sampling choice is already
representing a high k-point density in the reciprocal space, as large
periodic supercells are being used in the calculations.

In order to treat the PBE-GGA exchange and correlation potential
efficiently during the plane-wave calculations, we employ the approach
of White and Bird.\cite{white-bird}
In this method, by writing the energy and potential in a discretized
form, there is no need to represent the charge density and its
gradient on a finer real-space mesh.  This not only reduces the
computational time and memory requirements, but also improves
the convergence and stability of the calculations.

\section{The S\lowercase{i}/S\lowercase{i}O$_2$ interface}

The initial input structure for the MC annealing procedure is a
crystalline supercell containing either 12 or 40 layers of Si
atoms stacked along the $[001]$ direction, each layer consisting
of 4 atoms in a cell of dimensions $\sqrt{2}a\times\sqrt{2}a$ in
the $x$-$y$ plane.  The three central layers of Si atoms are held
fixed in each structure, to form the portion that will remain as
the c-Si region.  The bonds in that region, plus either one or
two layers of adjacent bonds, are initially assigned as S bonds;
the remainder are assigned as L bonds to form the ``oxide.'' (The
structure formed in this way is initially highly strained, but
the action of the MC bond-switching moves, combined with the fact
that the $c$-axis is allowed to relax on each MC step, soon
anneals away most of this strain.)  All bonds except those in the
central three-layer c-Si region are allowed to be switched and/or
converted during MC simulations. The MC annealing schedule starts with
a temperature on the order of roughly $10^4$K with five sweeps.
(Each
sweep is a number of trial steps equal to the total number of
atoms.) Then, the temperature is reduced by approximately a  factor of
two while the number of sweeps is increased by a factor of two
for each consecutive simulation stage. The annealing schedule stops
when the temperature reaches around 700K
and the number of sweeps reaches
the order of $10^2$.
Further reducing the temperature does not produce
a significant effect in the resulting structures.

In order to investigate the transition region near the interface
as generated by the MC procedure, we use initial input geometries
having 40 layers of Si atoms, having a total of 160 Si atoms and
256 O atoms. The initial dimensions of these periodic supercells are
7.68\AA\,$\times$ 7.68\AA\,$\times$ 54.30\AA,
but the third cell dimension relaxes to $\sim$90\AA\ during the
simulations.
The resulting statistical distribution of Si oxidation
number (i.e., the number of oxygen neighbors to a given Si
atom) {\it vs.} distance from the interface is shown in
\begin{figure}
\epsfxsize=2.8in
\centerline{\epsfbox{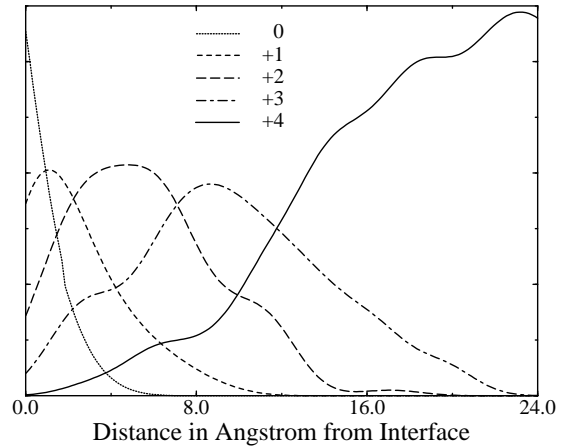}}
\vskip 0.5cm
\caption{Oxidation number of Si vs.\ distance of Si from
interface, as obtained from the Monte Carlo simulations.
Curves represent distribution profiles for five different oxidation
states, smoothed by a gaussian broadening of
width $\sim$2\AA. Data is compiled from six different
structures with a total of $\sim$$10^3$ Si atoms.}
\label{fig:oxidation}
\end{figure}
Fig.\ \ref{fig:oxidation}. It shows that the size of the transition
region is about 20 \AA, which agrees with some former experimental work.
\cite{gusev95,grunthaner86} The figure also shows that some Si$^{+1}$
species are concentrated within $\sim$5\AA\, of the interface,
while some Si$^{+2}$ species extend $\sim$10\AA\, into the a-SiO$_2$
region. These results agree very well with some XPS and PES
studies\cite{himpsel,grunthaner87} that find that Si$^{+1}$ and
Si$^{+2}$ states are localized within 6-10\AA\, of the interface.
These experiments also suggest that Si$^{+3}$ species extend
$\sim$30\AA\, into the bulk SiO$_2$, again in qualitative agreement
with our simulations.
Himpsel {\it et. al.}\cite{himpsel} have proposed that these characteristic
protrusions of Si$^{+3}$ into the bulk SiO$_2$ could be seen as the
cores of misfit dislocations, which are caused by the large lattice
mismatch of Si and SiO$_2$.

Of course, we emphasize that our statistical results reported
in Fig.\ \ref{fig:oxidation} are generated purely from the MC
simulations, with no input from the first-principles calculations.
The $z$-positions of the atoms would shift slightly
when relaxed with the first-principles calculations, but this
typically shifts the points in the figure by less than 0.2\AA.
More significantly, one should ideally recompute the statistics
using the first-principles total energies in the MC procedure,
but such an approach would be prohibitively expensive.
Nevertheless, the qualitative agreement with experiment regarding
the distribution of suboxide species suggests that the interface
structures as generated by the empirical MC procedure are fairly
realistic.

\begin{figure}
\epsfxsize=3.0in
\centerline{\epsfbox{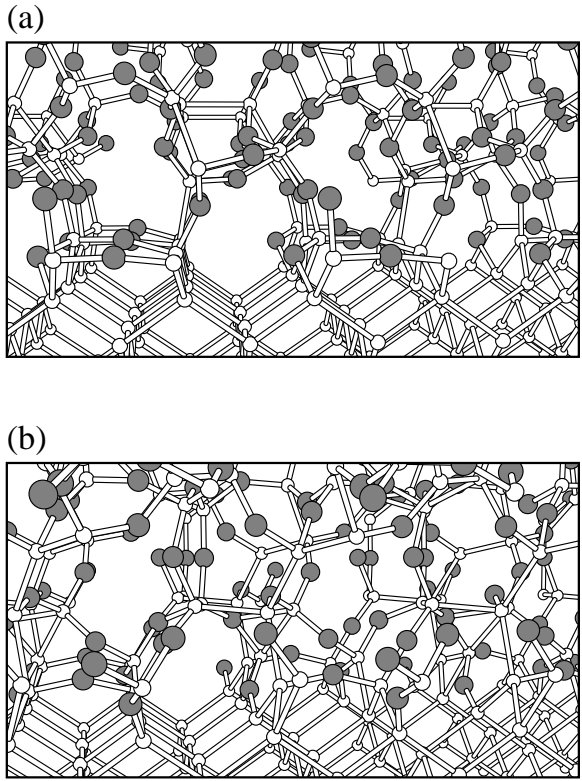}}
\vskip 0.5cm
\caption{Two examples of relaxed interface structures, as
generated by Monte Carlo annealing followed by first-principles
relaxation.  Periodic replicas arise because the field of view
spans two supercell dimensions in each of the directions parallel
to the interface.}
\label{fig:structure}
\end{figure}

First-principles calculations were only carried out for the
structures with 12 layers of Si atoms (a total of 96 or 112 atoms,
depending on how many layers were oxidized).  For the 40-layer
structures, such calculations would be too computationally demanding.
The initial input dimensions of the periodic supercell
are 7.68\AA\,$\times$ 7.68\AA\,$\times$ 16.29\AA, but this relaxes
to  7.68\AA\,$\times$ 7.68\AA\,$\times$ $\sim$23\AA\ when the third
dimension is allowed to relax during the MC procedure.
First-principles calculations are then carried out on seven structures
to relax both the atomic coordinates and the supercell $z$ dimension,
the latter typically being found to change by less than 6\% from the
empirical-potential MC value.
Note that these seven structures were generated using the same MC
procedure, but with different random number seeds, so that they can
be regarded as a small subset of the full ensemble that would be
generated by the MC approach.

\begin{figure}
\epsfxsize=2.4in
\centerline{\epsfbox{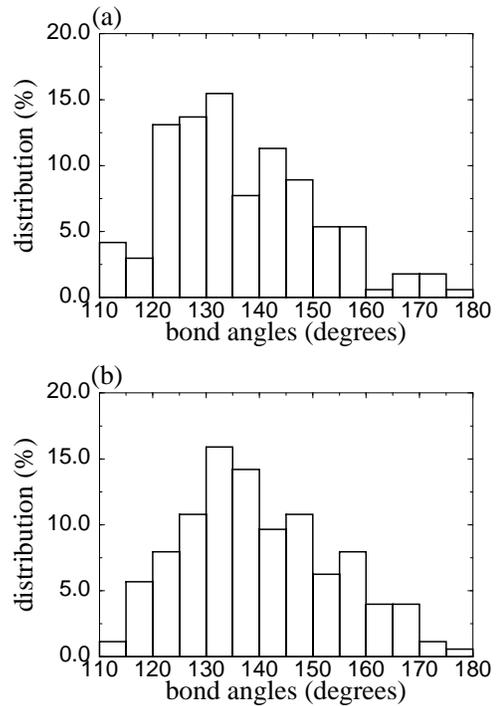}}
\vskip 0.5cm
\caption{Histograms of Si--O--Si bond angles, for bond angles
located (a) within 5 \AA\, of the interface, and (b)
between 5 and 10 \AA\, from the interface.}
\label{fig:angles}
\end{figure}

Two examples of the relaxed interface structures
are shown in Fig.\ \ref{fig:structure}.
Because of the lattice mismatch, a region of pure a-SiO$_2$ cannot
easily be attached directly to a c-Si substrate.  As a result,
as illustrated in the examples of Fig.\ \ref{fig:structure},
sub-oxide species are numerous in the region near the interface.
One might then expect that the structure in the oxide should
differ significantly from that of bulk a-SiO$_2$ in this
transition region.  Indeed, our results support this.  For example,
Fig.\ \ref{fig:angles} shows the statistical distribution of oxygen
bond angles in
different regions away from the interface. These distributions become
broader and more evenly distributed as one gets further from the
interface, eventually resembling fairly closely the accepted bulk
a-SiO$_2$ distribution.\cite{helms94}

Another structural aspect we have examined is the Si--Si bond length in the
SiO$_2$ region. We find that the bond lengths become shorter when the
average oxidation state of the two participating Si atoms increases
from zero to two. The bond length is about 4\% shorter than that of
c-Si (2.35\AA) when the average oxidation is two. As the average
oxidation state of the two Si increases to three, the bond length
rebounds slightly, but is still about 3\% shorter than that of the
c-Si.
Since an Si--O bond has less covalent character than a Si--Si bond,
the initial shortening of the bond can be understood as a strengthening
of the covalent character of the remaining Si--Si bonds, while
the rebound for Si$^{+3}$ is presumably related to the Coulomb
repulsion between the increasingly positively charged Si atoms as
they lose electrons to their more electronegative O neighbors.
The Si--O bond length decreases
from about 1.67\AA\, to about 1.62\AA\, when the
oxidation state of Si involved increases from one to four. As
expected, the Si$^{+4}$--O bond length is very close to the observed
value (1.6\AA) in a-SiO$_2$.

We have also investigated the electronic states of these
interface structures.  We do not find any defect state in the
energy gap of our structures. This is not surprising, and
reflects the fact that our geometries correspond to a continuous
random network structure without any broken bonds being present.
We also specifically inspected the highest occupied molecular
orbitals (HOMOs) and lowest unoccupied molecular orbitals (LUMOs)
of these structures, but did not find any simple pattern for the
localization of these orbitals.
Sometimes the HOMO or LUMO states are confined mainly to the c-Si
region (typically being concentrated either at the top or bottom
interface but remaining relatively delocalized in the $x$-$y$ plane),
but they also frequently have substantial amplitude at certain
localized sites in the oxide region.
We tried to determine whether abnormal bond lengths or bond
angles might be associated with the localization of the LUMO or
HOMO orbitals, but we found no simple correlation of this kind.

In summary, in this part of our work we have shown that it is
possible to generate reasonable interface structures for the
Si/SiO$_2$ interface despite the huge lattice mismatch, provided
that one allows for the presence of sub-oxide states of Si in the
transition region.  The distribution of sub-oxide states that
emerges from our simulations is generally consistent with
experiments.  Moreover, our interface structures are entirely
free of dangling bonds.  Of course, by its construction the MC
procedure automatically generates continuous random network (CRN)
structures that are free of dangling bonds, but we then relaxed
these structures using the first-principles calculations.  While
bonds might have broken or reformed during this relaxation
procedure, such events were never observed in our calculations.
That is, the first-principles relaxations always preserved the
identity of the MC-generated CRN structures.  This result is not
trivial, and provides strong support for the robustness of the MC
procedure.  The first-principles calculations also provided Si--O
bond-length and Si--O--Si bond-angle distributions that are
consistent with the the accepted picture for bulk a-SiO$_2$, and
the electronic spectrum was checked for the absence of gap
states.  For all these reasons, we argue that the two-step
procedure proposed here (MC generation followed by first-principles
relaxation) produces realistic interface structures.

\section{Oxidation kinetics in \lowercase{a}-S\lowercase{i}O$_2$}

In this part of the calculations, the
energetics of atomic oxygen in an a-SiO$_2$ environment is
investigated.  Thus, for the
MC simulations, only a random network of ``long'' bonds is needed
as input. Periodic supercells (10.10\AA $\times$ 10.10\AA\,$\times$
7.14\AA) containing 49 atoms (including the extra interstitial
O atom) are used.  The extra interstitial oxygen atom is first
placed randomly inside the cavity formed by the a-siO$_2$ network,
and the system is allowed to relax.  We find that the extra
oxygen atom binds to a neighboring oxygen atom that is already
part of a Si--O--Si bridge bond, as illustrated in Fig.~5(a).
The distance between the neighboring oxygens is only $\sim$1.54
to 1.59\AA, so that the bridge-bonded oxygen is essentially
threefold coordinated.

\begin{figure}
\epsfxsize=2.6in
\centerline{\epsfbox{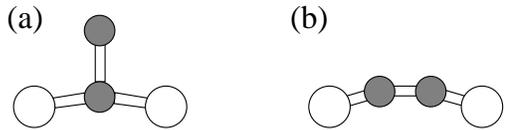}}
\vskip 0.5cm
\caption{Possible bonding configurations for an extra oxygen
atom in a-SiO$_2$.  (a) ``Threefold'' configuration in which an
extra oxygen binds to a bridge-bonded oxygen.
(b) ``Peroxyl bridge'' configuration in which an
extra oxygen inserts into a bridge bond.
Filled and open circles represent oxygen and
silicon atoms respectively.}
\label{fig:peroxyl}
\end{figure}

This threefold structure is a local minimum of
the energy landscape, but to determine whether it is truly stable
or only metastable, we then tried rearranging the two oxygen atoms
into a peroxyl bridge structure as shown in Fig.~5(b).  Again the
structure is allowed to relax, and a local energy minimum associated
with the peroxyl bridge configuration is found, with the distance
between the two oxygen atoms being $\sim$1.50 to 1.57\AA.
The existence of peroxyl bridge configurations in SiO$_2$
has already been suggested on the basis of some experimental
work\cite{nish89} in which UV absorption evidence for peroxyl bridges
was found in the presence of excess oxygen.
Comparing the energies of the threefold and peroxyl configurations,
we find that the peroxyl bridge is always the more stable configuration,
with the energy difference varying in the range of about
0.9 to 2.8eV.  We also find that there is a strong correlation
between the relative stability of the peroxyl configuration and
the length of the relaxed Si--Si distance in the peroxyl
bridge; thus, the peroxyl is (not surprisingly) most likely to form
where there is room for the two Si atoms to separate to make
room for it.

\begin{figure}
\epsfxsize=2.6in
\centerline{\epsfbox{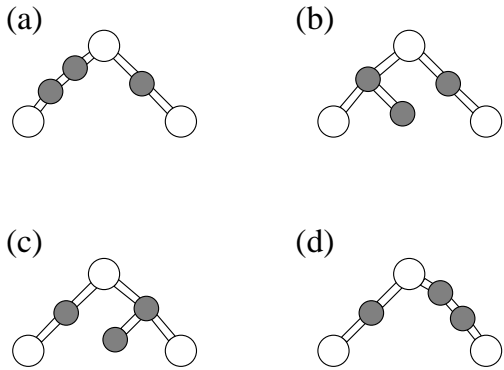}}
\vskip 0.5cm
\caption{Proposed mechanism for diffusion of atomic oxygen in a-SiO$_2$.
(a) Initial configuration with peroxyl bridge at left.
(b) Peroxyl bridge excites into a threefold structure.
(c) Extra oxygen atom jumps to a neighboring bond.
(d) Threefold structure decays to form new peroxyl bridge at right.}
\label{fig:hopping}
\end{figure}

We find almost no energy barrier between the threefold and
peroxyl configurations.  This suggests that the threefold
configuration is only barely metastable (i.e., almost unstable).
There could be two implications to be drawn from this
result. First, the peroxyl bridge configuration can assist in the
diffusion kinetics of oxygen. Spontaneous formation and disintegration
of peroxyl structures can occur during the oxidation process, and thus
helping in promoting the diffusion of oxygen. Second, the peroxyl
bridge structure can hop among neighboring bonds of Si. Starting from
the peroxyl bridge configuration, the hopping could occur by
an initial formation of
a threefold configuration, then a jump of the extra oxygen
atom to a neighboring bond, and then the formation of a new
peroxyl bridge in that bond.  This proposed hopping mechanism is
illustrated in Fig.\ \ref{fig:hopping}. The precise energy barrier
for this mechanism is not computed, as it depends in detail on
the specific environment. However, we expect it should be of the
same order of magnitude as the difference of the peroxyl bridge
and threefold configurations.

This hopping mechanism is similar in spirit to one proposed recently by
Hamann,\cite{hamann98}  who studied a peroxyl bridge structure
within an $\alpha$-quartz supercell. In Hamann's model, the overall hopping
process is similar to ours, except that in place of the intermediate
structure that we propose in Figs.\ \ref{fig:hopping}(b) and
\ref{fig:hopping}(c), he has proposed an intermediate structure
in which the extra oxygen atom is attached to the shared Si atom and
acts as its fifth neighbor. The process
then becomes a three-step mechanism. It begins with a peroxyl bridge,
then evolves into a fivefold Si configuration, and finally becomes a new
peroxyl bridge in another bond. Our energy difference between the
consecutive structures in the hopping mechanism is roughly similar to
that of Hamann's.  

\section{Summary}

We have studied a number of supercells to model the  Si(100)/a-SiO$_2$
interface. A Monte Carlo simulated-annealing method, applied to a
simple Keating-like model potential in which ``long'' and ``short''
bonds represent Si--O--Si and Si--Si bonds, is used to generate reasonable
structural models of the interface.  These structures are consistent
with experimental observations of sub-oxide distributions.
In particular, Si$^{+1}$ and Si$^{+2}$ sub-oxide states are
concentrated within about 10\AA\, of the interface, with
the Si$^{+3}$ extends up to about 20\AA\, into the bulk SiO$_2$. As
expected, the SiO$_2$ region shows more of the bulk a-SiO$_2$ character
as one moves away from the interface. We have also investigated the
HOMO and LUMO states of the interface structures.  We confirmed that
no deep gap states exist, but did not find any consistent pattern for
the nature of the HOMO and LUMO orbitals.

The kinetics of atomic oxygen in a-SiO$_2$ were also investigated.
The peroxyl bridge structure was found to be stable, with its energy
computed to be lower than that of interstitial of threefold
configurations.  We suggest a possible mechanism for transport
of atomic oxygen in the a-SiO$_2$ film, involving diffusion of
the peroxyl bridge configuration, that could play a role in the
interface oxidation during the formation of thin films. 

\acknowledgments
This work was supported by NSF grant DMR-96-13648.
Cray T90 supercomputer support was provided by an
NPACI allocation at SDSC.

\end{document}